\documentclass[12pt]{article}
\usepackage{amsfonts}
\usepackage{amsmath}
\usepackage{amssymb}
\usepackage{graphicx}

\begin{document}

\title{ \begin{flushright}
{\small CECS-PHY-06/18}
\end{flushright}
\vskip 1.0cm Three-dimensional supergravity reloaded}
\author{Alex Giacomini, Ricardo Troncoso and Steven Willison\thanks{%
E-mails: giacomini, steve, ratron-at-cecs.cl} \\
{\small Centro de Estudios Cient\'{\i}ficos (CECS), Casilla 1469, Valdivia,
Chile.}}
\date{}
\maketitle

\begin{abstract}
The locally supersymmetric extension of the most general gravity theory in
three dimensions leading to first order field equations for the vielbein and
the spin connection is constructed. Apart from the Einstein-Hilbert term
with cosmological constant, the gravitational sector contains the
Lorentz-Chern-Simons form and a term involving the torsion each with
arbitrary couplings. The supersymmetric extension is carried out for
vanishing and negative effective cosmological constant, and it is shown that
the action can be written as a Chern-Simons theory for the supersymmetric
extension of the Poincare and AdS groups, respectively. Here we introduce a
duality map between different gravity theories\ that greatly simplifies the
construction. This map relies on the different ways to make geometry emerge
from a single gauge potential. The extension for $\mathcal{N}=p+q$ gravitini
is also performed.
\end{abstract}

\section{Introduction}

Three dimensional gravity is a useful toy model to give an insight into the
problems of four dimensional gravity, which is a notoriously difficult
theory especially at the quantum level. Being a generally covariant theory
of gravity in three dimensions, it captures many of the basic conceptual
features of four dimensional General Relativity but avoiding many of the
computational difficulties.

The systematic study of three dimensional gravity was put forward by Deser,
Jackiw and 't Hooft \cite{Deser84}. The essence of the simplicity is due to
the fact that the gravitational field has no local degrees of freedom away
from the sources. In fact in vacuum the solutions are given by flat or
constant curvature spacetimes in the case of a cosmological constant. In
spite of this, the theory is not trivial as for example there are black hole
solutions (known as BTZ black holes) in the case of negative cosmological
constant \cite{BTZ}.

One of the most remarkable facts about three-dimensional gravity is that it
can be written as a gauge theory with Chern-Simons action \cite{townsend,
Witten}, where the gauge group is given by the AdS, dS or Poincar\'{e} group
for negative, positive and zero cosmological constant respectively.
Chern-Simons theory in three dimensions is a theory of flat connections.
This is related to the fact, mentioned above, that three dimensional gravity
has no local degrees of freedom. The locally supersymmetric extension of
General Relativity in three dimensions has been done by Deser and Kay~\cite%
{Deser-Kay}, and it has been shown that it can be written as a Chern-Simons
theory for negative \cite{townsend} or vanishing \cite{townsend2,HIPT}
cosmological constant.

To be more precise, the Chern-Simons gauge theory is equivalent to General
Relativity in the first order formalism. That is, the spin connection and
the vielbein are considered as independent variables\footnote{%
Conformal gravity and supergravity can also be formulated along these lines,
as done in~\cite{Horne-Witten} and~\cite{Lindstrom-Rocek} respectively.
\newline
Another nontrivial development of 3-D gravity is topologically massive
gravity \cite{Deser:1981wh}, where the Lorentz Chern-Simons term for the
spin connection is not independent, but a function of the vielbein. One
obtains then higher derivative field equations with massive gravitons. Also
in this theory black hole solutions are known and the thermodymamics has
been studied (see e.g., \cite{Solodukhin:2005ah} and references therein).
The supersymmetric extension of topologically massive gravity has been
studied in Ref. \cite{Deser-Kay}.}. However, there is a more general action
for gravity in three dimensions which apart from the Einstein-Hilbert term
with cosmological constant, contains the Lorentz-Chern-Simons form and a
term involving the torsion each with arbitrary couplings \cite%
{mielke,Mardones-Zanelli}. It has been shown that this model can also be
written as a Chern-Simons theory \cite{blagojevic,giacomini}. This is also
known as the Mielke-Baekler model. This is the three dimensional case of the
gravitational theories with torsion in arbitrary dimension considered in
Ref. \cite{Mardones-Zanelli}. This more general theory of gravity is
interesting also because it admits a black hole solution with negative
Riemannian curvature \cite{Garcia-Hehl} which is essentially a BTZ black
hole with torsion.

In this paper, we show how to construct the supersymmetric extension of this
more general model for negative and vanishing cosmological constant, first
for $\mathcal{N}=1$ and then for $\mathcal{N}$ gravitini. It is shown that
the construction is greatly simplified by making use of the mapping between
different theories of gravity presented here, which which allows one to map
between the standard theory without torsion and the more general theory
including torsion. This mapping is shown to arise as a consequence of
considering the different ways to make geometry emerge from a single gauge
potential. That is to say, there are different ways of relating the gauge
field of a Chern-Simons theory with the geometrical quantities, the vielbein
and spin connection. Hence, although the action is the same in terms of the
gauge field, the ambiguity in the identification of some of the gauge fields
with the vielbein actually generates different posibilities which turn out
to be physically inequivalent when one wants to make contact with gravity.
This becomes more explicit when one couples the theory to some standard
extenal source, which explicitly depends on the metric.

The structure of the paper is the following: In Section 2 we give a brief
review of the most general theory of gravity in first order formalism in
three dimensions. Then in Section 3 we construct its $\mathcal{N}=1$
supersymmetric extension. Section 4 is devoted to explore different ways of
making geometry to emerge from a gauge field. This turns out to be useful to
formulate a map between different theories of gravity, which is discussed in
Section 5. In Section 6 we show how the supergravity theory constructed in
Section 3 can be obtained applying the map discussed above to the standard
theory. Finally, in section 7 we show that the map is particularly useful in
order to perform the extension to $\mathcal{N}=p+q$ supersymmetries.

\section{ The most general first order action for three-dimensional gravity}

The most general gravitational action in three dimensions that is invariant
under local Lorentz transformations, up to a surface term, leading to first
order field equations for the dreibein $e^{a}$, and the spin connection $%
\omega ^{ab}$ is \cite{mielke,Mardones-Zanelli}
\begin{equation}
I_{G}(e^{a},\omega ^{a})=\int 2\alpha _{1}e^{a}R_{a}-\frac{\alpha _{2}}{3}%
\epsilon _{abc}e^{a}e^{b}e^{c}+\alpha _{3}L_{3}(\omega ^{a})+\alpha
_{4}e_{a}T^{a}\ .  \label{IG}
\end{equation}%
Here $\omega ^{a}$ stands for the dual of the spin connection, $\omega ^{a}:=%
\frac{1}{2}\epsilon ^{abc}\omega _{ab}$, and $L_{3}(\omega ^{a})$ is the
Lorentz-Chern-Simons form given by%
\begin{equation}
L_{3}(\omega ^{a})=\omega _{a}d\omega ^{a}+\frac{1}{3}\epsilon _{abc}\omega
^{a}\omega ^{b}\omega ^{c}\ .  \label{L3omega}
\end{equation}%
The wedge product between forms is understood.

The field equations obtained from the variation of (\ref{IG}) with respect
to the dreibein and the spin connection read
\begin{eqnarray}
2\alpha _{1}R_{a}-\alpha _{2}\epsilon _{abc}e^{b}e^{c}+2\alpha _{4}T_{a}
&=&0\ ,  \label{motion1} \\
2\alpha _{1}T_{a}+2\alpha _{3}R_{a}+\alpha _{4}\epsilon _{abc}e^{b}e^{c}
&=&0\ ,  \label{motion2}
\end{eqnarray}%
respectively, which imply that the three-dimensional geometry must be of
constant curvature and torsion\footnote{%
By constant torsion we mean that the torsion is constant and completely
antisymmetric when its indices are lowered.} given by
\begin{eqnarray}
R_{a} &=&\frac{1}{2}\left( \beta ^{2}+\frac{\sigma }{l^{2}}\right) \epsilon
_{abc}\ e^{b}e^{c}\ ,  \label{curvaure} \\
T_{a} &=&-\beta \epsilon _{abc}\ e^{b}e^{c}\ ,  \label{Torsion}
\end{eqnarray}%
where the constants $\beta $ and $l$ are fixed only by means of the
couplings appearing in the action\footnote{%
The case of $\alpha _{3}\alpha _{4}-\alpha _{1}^{2}=0$ is pathological,
since the field equations (\ref{motion1}) and (\ref{motion2}) become
linearly dependent.}
\begin{equation}
\beta =-\frac{1}{2}\frac{\alpha _{3}\alpha _{2}+\alpha _{4}\alpha _{1}}{%
\alpha _{3}\alpha _{4}-\alpha _{1}^{2}}\;\;\;\;;\;\;\;\;\;\beta ^{2}+\frac{%
\sigma }{l^{2}}=-\frac{\alpha _{1}\alpha _{2}+\alpha _{4}^{2}}{\alpha
_{3}\alpha _{4}-\alpha _{1}^{2}}\ .  \label{constants-AB}
\end{equation}%
We have introduced $\sigma =\pm 1$ or $0$.

The spin connection can be decomposed as
\begin{equation*}
\omega ^{ab}=\tilde{\omega}^{ab}+k^{ab}\ ,
\end{equation*}%
where $\tilde{\omega}^{ab}$ is the (torsion-free) Levi-Civita connection,
and $k^{ab}$ stands for the contorsion, which by Eq. (\ref{Torsion}) is
fixed as
\begin{equation}
k^{ab}=\beta \epsilon ^{abc}\ e_{c}\ .  \label{contorsion}
\end{equation}%
Thus, the curvature two-form can be decomposed in terms of its Riemannian
part $\tilde{R}^{ab}$ and the contorsion one-form $k^{ab}$ as%
\begin{equation*}
R^{ab}=\tilde{R}^{ab}+\tilde{D}k^{ab}+k_{c}^{a}k^{cb}\ ,
\end{equation*}%
where $\tilde{D}$ is the covariant derivative with the Levi-Civita
connection $\tilde{\omega}^{ab}$. Hence, Eqs. (\ref{curvaure}) and (\ref%
{contorsion}) imply that spacetime has constant curvature since the
Riemannian part of the curvature acquires the form%
\begin{equation}
\tilde{R}^{ab}=\frac{\sigma }{l^{2}}\ e^{a}e^{b}\ .
\label{constant-curvature}
\end{equation}%
where $\frac{\sigma }{l^{2}}$ is the effective cosmological constant.

The minimal local supersymmetric extension of the action (\ref{IG}) is
carried out in the next section.

\section{ Locally supersymmetric extension with\newline
$\mathcal{N}=1$}

Let us consider the following action principle%
\begin{equation}
I(e^{a},\omega ^{a},\psi )=I_{G}+I_{\psi }\ ,  \label{Isusy1}
\end{equation}%
where the gravitational sector is given by $I_{G}$ in Eq. (\ref{IG}), and
the fermionic term reads
\begin{equation}
I_{\psi }=-\left( \alpha _{1}-\alpha _{3}\left( \beta -\frac{1}{l}\right)
\right) \ \bar{\psi}\left( D-\frac{1}{2}\left( \beta +\frac{1}{l}\right)
e^{a}\Gamma _{a}\right) \psi \ ,
\end{equation}%
where $\psi =\psi _{\mu }dx^{\mu }$ is the gravitino, whose Lorentz
covariant derivative is given by $D\psi =d\psi -\frac{1}{2}\omega ^{a}\Gamma
_{a}$ . Therefore, when the effective cosmological constant is negative or
zero one can see that, the action (\ref{Isusy1}) is invariant, up to a
surface term, under the following local supersymmetry transformations
spanned by $\epsilon =\epsilon (x)$,
\begin{gather}
\delta e^{a}=\frac{1}{2}\bar{\epsilon}\Gamma ^{a}\psi \ ,  \notag \\
\delta \omega ^{a}=-\frac{1}{2}\left( \beta -\frac{1}{l}\right) \bar{\epsilon%
}\Gamma ^{a}\psi \ ,  \label{delta-e+omega+psi-1} \\
\delta \psi =D\epsilon -\frac{1}{2}\left( \beta +\frac{1}{l}\right)
e^{a}\Gamma _{a}\epsilon \ .  \notag
\end{gather}%
Note that apart from diffeormorphisms, local supersymmetry transformations (%
\ref{delta-e+omega+psi-1}), and the manifest local Lorentz symmetry spanned
by, $\lambda ^{a}=\lambda ^{a}(x)$,%
\begin{gather}
\delta e^{a}=\epsilon ^{abc}e_{b}\lambda _{c}\ ,  \notag \\
\delta \omega ^{a}=D\lambda ^{a}\ ,  \label{Lorentz} \\
\delta \psi =-\frac{1}{2}\lambda ^{a}\Gamma _{a}\psi \ ,  \notag
\end{gather}%
the action (\ref{Isusy1}) has an extra local symmetry which is spanned by a
vector $\chi ^{a}=\chi ^{a}(x)$:%
\begin{gather}
\delta e^{a}=D\chi ^{a}+\beta \epsilon ^{abc}e_{b}\chi _{c},  \notag \\
\delta \omega ^{a}=\frac{1}{l^{2}}\epsilon ^{abc}e_{b}\chi _{c}-\beta \left(
D\chi ^{a}+\beta \epsilon ^{abc}e^{b}\chi ^{a}\right) ,  \label{AdS-boost} \\
\delta \psi =\frac{1}{2l}\chi ^{a}\Gamma _{a}\psi \ .  \notag
\end{gather}%
One may check that the transformations (\ref{AdS-boost}) together with the
local Lorentz symmetry (\ref{Lorentz}) close, merging into the AdS algebra $%
SO(2,2)$ which is isomorphic to $Sp(2)\times Sp(2)$. Once the supersymmetry
tranformations are included, one verifies that the whole set in Eqs. (\ref%
{delta-e+omega+psi-1}), (\ref{Lorentz}), and (\ref{AdS-boost}) closes
off-shell and without the need of auxiliary fields into the $OSp(2|1)\times
Sp(2)$ superalgebra, which is the minimal supersymmetric extension of the
AdS algebra.

Analogously, in the limit $l\rightarrow \infty $, the transformations (\ref%
{delta-e+omega+psi-1}), (\ref{Lorentz}), and (\ref{AdS-boost}) close
off-shell in the Super Poincar\'{e} algebra with $\mathcal{N}=1$.

It is natural then to see whether the action principle (\ref{Isusy1}) can be
written in a manifestly covariant way under this gauge symmetry. It is worth
pointing out that the local symmetries (\ref{delta-e+omega+psi-1}), (\ref%
{Lorentz}), and (\ref{AdS-boost}) can be obtained from gauge tranformations
of a single super AdS connection given by%
\begin{equation}
A=\left( \omega ^{a}+\beta e^{a}\right) J_{a}+\frac{e^{a}}{l}P_{a}+\frac{1}{%
\sqrt{l}}\psi ^{\alpha }Q_{\alpha }\ ,  \label{A}
\end{equation}%
where $J_{a}$, $P_{a}$, and $Q_{\alpha }$ are the generators of Lorentz
transformations, AdS boosts, and supersymmetry, respectively, spanning the $%
OSp(2|1)\times Sp(2)$\ algebra\footnote{%
The relative sign at the r.h.s. of the anticommutator corresponds to which
of the copies of $Sp(2)$ has been extended to $OSp(2|1)$. Choosing the other
copy amounts to make the replacement $l\rightarrow -l$ everywhere, and
reversing the overall sign in the fermionic action.
\par
In the vanishing cosmological constant limit the super Poincar\'{e} algebra
is recovered taking the In\"{o}n\"{u}-Wigner contraction. This is performed
through rescaling the generators as $P_{a}\rightarrow l^{-1}P_{a}$, $%
Q_{\alpha }\rightarrow l^{-1/2}Q_{\alpha }$, and taking the limit $%
l\rightarrow \infty $.}:%
\begin{gather}
\lbrack J_{a},J_{b}]=\epsilon _{abc}J^{c}\;\;\;;\;\;\;[J_{a},P_{b}]=\epsilon
_{abc}P^{c}\;\;\;;\;\;\;[P_{a},P_{b}]=\epsilon _{abc}J^{c}\ ,
\label{bosonic_Lie} \\
\lbrack J_{a},Q_{\alpha }]=[P_{a},Q_{\alpha }]=-\frac{1}{2}(\Gamma
_{a})_{\alpha }^{\beta }Q_{\beta }\quad ;\quad \{Q_{\alpha },Q_{\beta
}\}=(C\Gamma ^{a})_{\alpha \beta }(J_{a}+P_{a}).  \label{super_AdS}
\end{gather}

Indeed, for a gauge transformation
\begin{equation}
\delta A=d\lambda +\left[ A,\lambda \right] ,  \label{Gauge
transf}
\end{equation}%
choosing the Lie algebra valued parameter as $\lambda =\lambda ^{a}J_{a}$
the Lorentz transformations (\ref{Lorentz}) are recovered, and choosing $%
\lambda =\frac{\chi ^{a}}{l}P_{a}$ one obtains the AdS boosts in Eq. (\ref%
{AdS-boost}). Local supersymmetry transformations (\ref{delta-e+omega+psi-1}%
) are similarly obtained from (\ref{Gauge transf}) using $\lambda =\epsilon
^{\alpha }Q_{\alpha }$.

Therefore, the dreibein, the spin connection and the gravitino can be seen
as different components of a single connection $A$ for the gauge group $%
OSp(2|1)\times Sp(2)$ as in Eq. (\ref{A}). Using the fact that the gauge
group admits two independent bilinear invariant forms, one can show that the
supergravity action (\ref{Isusy1}) can be written, up to a surface term, as
a Chern-Simons theory for the gauge field (\ref{A}). This task is explicitly
performed in section \ref{N1_section} and the extension for $\mathcal{N}=p+q$
gravitini is done in section \ref{pq_section}.

Note that the identification of the components of the gauge potential in
equation (\ref{A}) with the vielbein and the spin connection is not
performed in the standard way as in Refs. \cite{townsend}, \cite{Witten}.
For the purely gravitational sector, this general identification is shown to
be equivalent to the one performed in \cite{blagojevic} for the AdS group,
as well as the one in \cite{giacomini} for the dS and Poincar\'{e}. This
fact was used to show that the gravitational sector can be formulated as a
Chern-Simons theory for any value of the effective cosmological constant $%
\sigma/l^2$, which determines the gauge group.

\section{Different ways to make geometry emerge from a single gauge potential%
}

Here we address the issue of different possibilities to make geometry emerge
from a connection for the AdS, dS and Poincare groups.

Any component of the gauge potential can not be identified arbitrarily with
the vielbein and the spin connection since in general they would not
transform suitably under the corresponding Lorentz subgroup. Thus the
possible identifications must satisfy the requirement that under a local
Lorentz transformation with parameter $\lambda ^{a}$ they transform as

\begin{gather}
\delta e^{a}=\epsilon ^{abc}e_{b}\lambda _{c}\ ,  \label{deltae} \\
\delta \omega ^{a}=D\lambda ^{a}\ .  \label{deltaomega}
\end{gather}

Let us explicitly find different possible identifications between gauge
fields for the AdS group and geometry in three dimensions. The AdS group is
spanned by the Lorentz generators $J_{a}$ and the AdS boosts $P_{a}$ with
the Lie algebra (\ref{bosonic_Lie}). A gauge connection for the AdS group
then reads
\begin{equation}
A=B^{a}J_{a}+C^{a}P_{a}\ ,
\end{equation}%
where $B^{a}$ and $C^{a}$ are its components. Let us consider a Lie
algebra-valued parameter of the form%
\begin{equation}
\lambda =\lambda ^{a}J_{a}+\chi ^{a}P_{a}\ ,
\end{equation}%
Thus, under an infinitesimal gauge transformation the connection transforms
as
\begin{equation*}
\delta A=d\lambda +[A,\lambda ]\ ,
\end{equation*}%
which means that its components transform according to%
\begin{eqnarray}
\delta B^{a} &=&d\lambda ^{a}+B^{a}\lambda ^{b}\epsilon _{abc}+C^{a}\chi
^{b}\epsilon _{abc}\ ,  \label{B} \\
\delta C^{a} &=&d\chi ^{a}+C^{a}\lambda ^{b}\epsilon _{abc}+B^{a}\chi
^{b}\epsilon _{abc}\ .  \label{C}
\end{eqnarray}%
Now in order to relate the spin connection and dreibein with the gauge
fields $B^{a}$ and $C^{a}$, it is natural to consider an arbitrary
combination of them. We must verify that they transform as in Eqs. (\ref%
{deltae}) and (\ref{deltaomega}) under a local Lorentz subgroup.

For simplicity let us consider a linear relation between the gauge fields $%
B^{a}$, $C^{a}$ with the vielbein $e^{a}$ and spin connection $\omega ^{a}$
of the form\footnote{%
Note this is not the most general linear combinations which can be
considered. An example that is not contained in our ansatz is $%
A=-e^{0}P_{0}+P_{i}\epsilon ^{0ij}\omega _{j}+\omega ^{0}J_{0}+J_{i}\epsilon
^{0ij}e_{j}$. For this choice the dreibein and the spin connection transform
in the correct way under the Lorentz group. This possibility arises from the
fact that the AdS group is $SO(2,2)$, so there is an ambiguity in choosing
the time direction. Notice that this is just an AdS transformation and so it
does not change the form of the action.}
\begin{eqnarray}
B^{a} &=&\alpha \omega ^{a}+\beta e^{a}\ , \\
C^{a} &=&\gamma e^{a}+\mu \omega ^{a}\ .
\end{eqnarray}%
Inserting these last equations in (\ref{B}), (\ref{C}) we obtain the
transformation law for the would-be vielbein and spin connection $e^{a}$ and
$\omega ^{a}$. For the would-be vielbein one obtains that it transforms as%
\begin{eqnarray}
(\mu \beta -\gamma \alpha )\delta e_{c} &=&d(\mu \lambda _{c}-\alpha \chi
_{c})+  \notag \\
&&\mu (\alpha \omega ^{a}\lambda ^{b}\epsilon _{abc}+\beta e^{a}\lambda
^{b}\epsilon _{abc}+\gamma e^{a}\chi ^{a}\epsilon _{abc}+\mu \omega ^{a}\chi
^{b}\epsilon _{abc})  \label{deltae_general} \\
&&-\alpha (\gamma e^{a}\lambda ^{b}\epsilon _{abc}+\mu \omega ^{a}\lambda
^{b}\epsilon _{abc}+\alpha \omega ^{a}\chi ^{b}\epsilon _{abc}+\beta
e^{a}\chi ^{b}\epsilon _{abc})\ .  \notag
\end{eqnarray}%
Since the vielbein must transform as a vector under the Lorentz group as
given in eq.(\ref{deltae}), the inhomogeneous term containing the exterior
derivative in Eq. (\ref{deltae_general}) necessarily has to do with AdS
boosts, as it occurs in Eq. (\ref{AdS-boost}), with parameter $\alpha \chi
^{a}-\mu \lambda ^{a}$. As we need to focus just on the local Lorentz
tranformations, without loosing generality, we can impose the AdS boost
parameter to vanish, i.e.,%
\begin{equation*}
\alpha \chi ^{a}=\mu \lambda ^{a}\ .
\end{equation*}%
One can then insert this back in (\ref{deltae_general}), but in order to
recover the correct tranformation law for the vielbein, one must impose the
cancelation of the terms containing $\omega ^{a}$, which leads to the
condition%
\begin{equation*}
\mu \left( \alpha ^{2}-\mu ^{2}\right) =0\ ,
\end{equation*}%
which is solved for $\mu =0$ or $\mu ^{2}=\alpha ^{2}$. The case $\mu
^{2}=\alpha ^{2}$ is discarded since it would transform as a scalar, $\delta
e^{a}=0$, so that $e^{a}$ cannot be identified with the vielbein. The
remaining possibility, $\mu =0$, exactly reproduces the vector
transformation law (\ref{deltae}) required for the vielbein.

The transformation law for the would-be spin connection $\omega ^{a}$ can
then be found plugging the above results back in Eq. (\ref{B}), and it reads%
\begin{equation*}
\alpha \delta \omega ^{a}=d\lambda ^{a}+\alpha \epsilon ^{abc}\omega
_{b}\lambda _{c}\ .
\end{equation*}%
Hence, $\alpha =1$ is required to reproduce the transformation (\ref%
{deltaomega}), so that $\omega ^{a}$ can be identified with the spin
connection.

In sum, according to our ansatz, the way to make geometry emerge from a
connection for the AdS group is obtained extracting the spin connection and
the vielbein from the gauge fields $B^{a}$ and $C^{a}$ from the following
relation%
\begin{eqnarray}
B^{a} &=&\omega ^{a}+\beta e^{a}\ ,  \notag \\
C^{a} &=&\frac{1}{l}e^{a}\ ,  \label{Identification}
\end{eqnarray}%
which depends on two arbitrary parameters $\beta $ and $l:=\gamma ^{-1}$, so
that the gauge field is given by%
\begin{equation}
A=(\omega ^{a}+\beta e^{a})J_{a}+\frac{1}{l}e^{a}P_{a}\ .  \label{Agrav}
\end{equation}

Following the same reasoning, the same identification as in Eq. (\ref%
{Identification}), or equivalently (\ref{Agrav}), can be readily found for
the Poincar\'{e} and dS groups. For all the cases the algebra reads
\begin{equation}
\lbrack J_{a},J_{b}]=\epsilon _{abc}J^{c}\;\;\;;\;\;\;[J_{a},P_{b}]=\epsilon
_{abc}P^{c}\;\;\;;\;\;\;[P_{a},P_{b}]=-\sigma \epsilon _{abc}J^{c}\ ,
\end{equation}%
where
\begin{equation*}
\sigma =\left\{
\begin{array}{ccc}
1 & : & \mathrm{dS} \\
0 & : & \mathrm{Poincar\acute{e}} \\
-1 & : & \mathrm{AdS}%
\end{array}%
\right. \ .
\end{equation*}%
It can be seen that the parameters $l$ and $\beta $ acquire a precise
geometrical interpretation when one deals with flat connections as follows.
The identification (\ref{Identification}) leads to the curvature $F=dA+A^{2}$%
, which in components reads%
\begin{equation*}
F=\left( R^{a}+\beta T^{a}+\frac{1}{2}\left( \beta ^{2}-\frac{\sigma }{l^{2}}%
\right) \epsilon ^{abc}e_{b}e_{c}\text{ }\right) J_{a}+\frac{1}{l}\left(
T^{a}+\beta \epsilon ^{abc}e_{b}e_{c}\right) \text{ }P_{a}\ ,
\end{equation*}%
so that for flat connections, characterized by $F=0$, the curvature and
torsion two-forms are given by%
\begin{eqnarray}
R_{a} &=&\frac{1}{2}\left( \beta ^{2}+\frac{\sigma }{l^{2}}\right) \epsilon
_{abc}e^{b}e^{c}\ ,  \label{Ra} \\
T_{a} &=&-\beta \epsilon _{abc}e^{b}e^{c}\ ,  \label{Ta}
\end{eqnarray}%
so that the geometry is decribed by spacetimes of constant curvature and
torsion. Therefore, $\beta $ parametrizes the torsion and $l$ parametrizes
the radius of curvature, since its Riemannian part turns out to be
\begin{equation*}
\tilde{R}^{ab}=\frac{\sigma }{l^{2}}e^{a}e^{b}\ .
\end{equation*}%
Note that since Eqs. (\ref{Ra}) and (\ref{Ta}) were obtained from requiring
the gauge field (\ref{Agrav}) to be locally flat, i.e., from $F=0$, they
correspond to the field equations for a Chern-Simons theory. On the other
hand, since Eqs. (\ref{Ra}) and (\ref{Ta}) are exactly the same as the ones
coming from the most general action for gravity in three-dimensions
described by (\ref{IG}) it goes without saying that, at least at the level
of the field equations, both theories are equivalent.

The standard identification done for torsionless gravity, as in Refs. \cite%
{townsend,Witten}, corresponds then to setting $\beta =0$.

\section{A map between different geometries and gravity theories}

\label{Map}

Here we show that the identification that allows one to extract the vielbein
and the spin connection from a gauge potential given by Eq. (\ref%
{Identification}) naturally provides a map between different geometries
which are solutions of the Mielke-Baekler model with different coupling
constants.

This can be seen as follows: suppose we have an AdS, dS or Poincar\'{e}
gauge theory with connection $A$. Starting from the conventional
identification $\beta =0$ and $\gamma =\frac{1}{l}$, we can define the
following mapping:
\begin{equation}
\omega ^{a}\rightarrow \omega ^{a}+\beta e^{a};\quad e^{a}\rightarrow e^{a},
\label{shift}
\end{equation}%
i.e. the spin connection is shifted by a term proportional to the vielbein.
Because of the results of the previous section, this mapping does not spoil
the identification between gauge fields and geometry. But it changes the
form of the action when written in terms of the spin connection and the
vielbein. Note that we could also have rescaled the vielbein. However this
would just amount to a rescaling of the cosmological constant and so can be
neglected without loss of generality.

For example, let us see the effect of this mapping on the AdS action. The
most general action for gravity with negative cosmological constant giving
field equations that imply vanishing torsion in three dimensions, is given
by the sum of the standard Einstein-Hilbert and the exotic action. This can
be written as:
\begin{equation}
I_{AdS}=\int \kappa \left( 2R^{a}e_{a}+\frac{1}{3l^{2}}\epsilon
_{abc}e^{a}e^{b}e^{c}\right) +k\left( L_{3}(\omega ^{a})+\frac{1}{l^{2}}%
T^{a}e_{a}\right) .  \label{ADS_notorsion}
\end{equation}%
The existence of an exotic action is due to the fact that for the AdS
algebra in three dimensions we have two nondegenerate invariant bilinear
forms. This is not the most general gravity theory in three dimensions since
it has only three free parameters. This action gives as equations of motion
\begin{equation*}
2R^{a}=-\frac{1}{l^{2}}\epsilon _{bc}^{a}e^{b}e^{c}\;\;\;;\;\;\;T^{a}=0
\end{equation*}%
Making now the shift in the spin connection as defined in equation (\ref%
{shift}) the action (\ref{ADS_notorsion}) is mapped into the following one:
\begin{equation}
I^{\prime }=\int \left( 2{\kappa }+2k\beta \right) R^{a}e_{a}+\left( \frac{{%
\kappa }}{3l^{2}}+{\kappa }\beta ^{2}+\frac{k}{l^{2}}\beta +k\frac{\beta ^{3}%
}{3}\right) \epsilon _{abc}e^{a}e^{b}e^{c}
\end{equation}%
\begin{equation*}
+kL_{3}(\omega )+\left( \frac{k}{l^{2}}+2{\kappa }\beta +k\beta ^{2}\right)
T^{a}e_{a}
\end{equation*}

The starting action (\ref{ADS_notorsion}) had three free parameters namely $%
\kappa $, $k$ and $1/l^{2}$ but now we have four free parameters, $\beta $
being arbitrary. The action $I^{\prime }$ is therefore is of the form given
in (\ref{IG}) with coefficients
\begin{gather}
\alpha _{1}={\kappa }+k\beta , \\
\alpha _{2}=-3\left( \frac{{\kappa }}{3l^{2}}+{\kappa }\beta ^{2}+\frac{k}{%
l^{2}}\beta +k\frac{\beta ^{3}}{3}\right) , \\
\alpha _{3}=k, \\
\alpha _{4}=\frac{k}{l^{2}}+2{\kappa }\beta +k\beta ^{2}.  \label{alpha}
\end{gather}

The shift in the spin connection has therefore the effect of mapping the
action (\ref{ADS_notorsion}) into a Mielke-Baekler action (\ref{IG}) for
every value of $\beta $. Furthermore we can invert the relations (\ref{alpha}%
)
\begin{gather}
k=\alpha _{3}, \\
\kappa =\alpha _{1}-\alpha _{3}\beta ,
\end{gather}%
and $\beta $ and $l$ are given by (\ref{constants-AB}). This means that,
starting from the action (\ref{ADS_notorsion}), with a suitable choice of $%
\beta $ we can recover every point in the parameter space of the
Mielke-Baekler model which has negative effective cosmological constant.
Therefore also the solutions of the original action (\ref{ADS_notorsion})
are mapped into solutions of the Mielke Baekler model. We would like to
stress that this fact can be used to greatly simplify the construction of
the supersymmetric extension of the Mielke Baekler model, since this task
can be done by knowing the SUSY extension of the action (\ref{ADS_notorsion}%
), and applying the map.

The parameter $\beta $ which characterises the field redefinition, is
related to the contorsion of the resulting theory by (\ref{contorsion}). The
effective cosmological constant $1/l^{2}$ is left unchanged by the field
redefinition. Note that the map presented here can also be applyed in the
cases where the effective cosmological constant is nonnegative.

As an application of the map, in the next section we show how the
supergravity theory constructed in section 3 can be recovered from the
standard supergravity theory with $\mathcal{N}=1$.

\section{Recovering the most general supergravity theory with $\mathcal{N}=1$
from the map}

\label{N1_section}

It is useful to recall that in three dimensions, the AdS group is isomorphic
to $Sp(2)\times Sp(2)$. This can be easily seen changing the basis of the
AdS algebra ( \ref{bosonic_Lie}) according to $J_{a}=J_{a}^{+}+J_{a}^{-}$
and $P_{a}=J_{a}^{+}-J_{a}^{-}$, where $J_{a}^{+}$ and $J_{a}^{-}$ are the
generators of each individual copy of $Sp(2)$. The connection in this basis
is given by
\begin{equation*}
A=A_{+}^{a}J_{a}^{+}+A_{-}^{a}J_{a}^{-}.
\end{equation*}%
The standard way of making geometry emerge, in this basis reads:
\begin{equation*}
A_{+}^{a}=\omega ^{a}+\frac{1}{l}e^{a}\,,\qquad A_{-}^{a}=\omega ^{a}-\frac{1%
}{l}e^{a}.
\end{equation*}
The most general AdS gravity theory, obtained by making the shift (\ref%
{shift}), in this basis is:
\begin{equation*}
A_{+}^{a}=\omega ^{a}+\left( \beta +\frac{1}{l}\right) e^{a}\,,\qquad
A_{-}^{a}=\omega ^{a}+\left( \beta -\frac{1}{l}\right) e^{a}\,.
\end{equation*}

This basis is particularly suitable for treating supergravity because the
minimal supersymmetric extension of the AdS algebra is obtained just by
supersymmetrizing one of the copies, that is $OSp(2|1)\times Sp(2)$. The $%
OSp(2|1)$ algebra is:
\begin{align}
\lbrack J_{a}^{+},J_{b}^{+}]& =\epsilon _{abc}J_{+}^{c},  \notag \\
\lbrack J_{a}^{+},Q_{\alpha }]& =-\frac{1}{2}(\Gamma _{a})_{\alpha }^{\beta
}Q_{\beta },  \label{osp1} \\
\{Q_{\alpha },Q_{\beta }\}& =(C\Gamma ^{a})_{\alpha \beta }J_{a}^{+}.  \notag
\end{align}
The supersymmetric AdS action will be made from a $OSp(2|1)$ connection $%
\mathcal{A}_{+}=A_{+}^{a}J_{a}^{+}+\psi ^{\alpha }Q_{\alpha }/\sqrt{l}$ and
a $Sp(2)$ connection $\mathcal{A}_{-}=A_{-}^{a}J_{a}^{-}$.

Let us first review the supersymmetric Einstein-Hilbert action with negative
cosmological constant. This is constructed as a Chern-Simons theory whose
Lagrangian, $L,$ satisfies $\left\langle \mathcal{F}_{+}^{2}\right\rangle
-\left\langle \mathcal{F}_{-}^{2}\right\rangle =dL$, making the standard
identification $A_{+}^{a}=\omega ^{a}+e^{a}/l$ and $A_{-}^{a}=\omega
^{a}-e^{a}/l$. We use the matrix representation of the generators given in
Appendix \ref{Appendix_N1}. In this case the invariant bilinear form is the
(super)trace. The action reads:
\begin{equation}
I_{1}=\hat{\kappa}\int_{M}\frac{1}{2}\left\{ L_{\text{cs}}(A_{+})-L_{\text{CS%
}}\ (A_{-})\right\} -\frac{1}{l}\bar{\psi}\nabla \psi
\end{equation}%
where $L_{\text{CS}}(A)=\langle AdA+\frac{2}{3}A^{3} \rangle$. The action
reduces to

\begin{equation*}
I_{1}=\frac{\hat{\kappa}}{l}\int_{M}2e^{a}R_{a}+\frac{1}{3l^{2}}\epsilon
_{abc}e^{a}e^{b}e^{c}-\bar{\psi}\nabla \psi ,
\end{equation*}%
up to a surface term. Here $\nabla \psi $ is the covariant derivative with
respect to the connection $A_{+}$, which, in terms of the vielbein and spin
connection is $\nabla \psi =(d\psi -\frac{1}{2}\omega ^{a}\Gamma _{a}\psi -%
\frac{1}{2l}e^{a}\Gamma _{a}\psi )\equiv D\psi -\frac{1}{2l}e^{a}\Gamma
_{a}\psi $.

In order to obtain the most general supersymmetric AdS action in three
dimensions with $\mathcal{N}=1$, we proceed as follows: we add to the
standard supersymmetric Einstein-Hilbert action the supersymmetric
\textquotedblleft exotic action". The exotic action is constructed from $%
\mathrm{STr}(\mathcal{F}_{+}^{2})+\mathrm{Tr}(\mathcal{F}_{-}^{2})=dL_{\text{%
exotic}}$, and it reads
\begin{equation*}
I_{\text{exotic}}\ ={k}\int_{M}\frac{1}{l^{2}}e^{a}T_{a}+I_{\text{CS}%
}(\omega )-\frac{1}{l}\bar{\psi}\nabla \psi .
\end{equation*}
The sum of the two actions $I_{1}$ and $I_{\text{exotic}}$ reads:
\begin{equation*}
I = \int_{M}\kappa \left( 2e^{a}R_{a}+\frac{1}{3l^{2}}\epsilon
_{abc}e^{a}e^{b}e^{c}\right) +{k}\left( I_{\text{CS}}(\omega )+\frac{1}{l^{2}%
}e^{a}T_{a}\right) -\left( \kappa +\frac{k}{l}\right) \bar{\psi}\nabla \psi
\end{equation*}%
where we have redefined $\kappa =\hat{\kappa}/l$.

The last action is the starting point for recovering the $\mathcal{N}=1$
supersymmetric extension of the most general gravity theory, described by
the action (\ref{Isusy1}). In fact applying the mapping $\omega
^{a}\rightarrow \omega ^{a}+\beta e^{a}$ \ the previous action is mapped
\begin{equation}
I\rightarrow I^{\prime }
\end{equation}%
where $I^{\prime }$ is given by
\begin{equation}
I^{\prime }=\left( 2\kappa +2k\beta \right) R^{a}e_{a}+\left( \frac{\kappa }{%
3l^{2}}+\kappa \beta ^{2}+\frac{k}{l^{2}}\beta +k\frac{\beta ^{3}}{3}\right)
\epsilon _{abc}e^{a}e^{b}e^{c}
\end{equation}%
\begin{equation*}
+kI_{CS}+\left( \frac{k}{l^{2}}+2\kappa \beta +k\beta ^{2}\right)
T^{a}e_{a}-\left( \kappa +\frac{k}{l}\right) \bar{\psi}(\nabla -\frac{\beta
}{2}e^{a}\Gamma _{a})\psi \ .
\end{equation*}%
We make now the same redefinitions as in (\ref{alpha}) and so we obtain the
action $I_{G}$ of eq. (\ref{IG}) plus the following fermionic term
\begin{equation*}
-\left( \alpha _{1}-\alpha _{3}\left( \beta -\frac{1}{l}\right) \right) \bar{%
\psi}\left( D-\frac{1}{2}\left( \beta +\frac{1}{l}\right) e^{a}\Gamma
_{a}\right) \psi
\end{equation*}%
And the supersymmetry transformations can be obtained by applying the shift (%
\ref{shift}) to the known transformations of the torsion free case which are
then as in (\ref{delta-e+omega+psi-1}). The Poincare case is obtained for $%
\mathcal{N}=1$ simply by taking the limit $l\rightarrow \infty $.

Notice that for $\beta^2 -1/l^2 =0$ the vielbein decouples from the
fermionic term and so the gravitino does not contribute to the stress-energy
tensor. This is related to the fact that in this case one of the $SL(2,R)$
connections becomes $A_+ = \omega ^a J_a$. This case is characterized by
zero curvature but nonzero torsion, which in literature is known as
teleparallel gravity (as a review see e.g. \cite{teleparallel}).

\section{The $\mathcal{N}=p+q$ gravitini case}

\label{pq_section}

In this section the locally supersymmetric extension of the most general
theory of gravity in three dimensions for $\mathcal{N}=p+q$ gravitini is
constructed. In order to do this we will start from the linear combination
of the $\mathcal{N}=p+q$ action of standard three-dimensional supergravity
of Achucarro and Townsend \cite{townsend}, and the $\mathcal{N}=p+q$
\textquotedblleft exotic" theory, and then apply the map in order to obtain
the local supersymmetric extension of the more general theory. Let us
therefore briefly recall the model of Achucarro and Townsend. As the AdS
group is isomorphic to the $Sp(2)\times Sp(2)$ group we can obtain a
supersymmetric extension with $\mathcal{N}$ gravitini of the AdS theory by
supersymmetrizing separately each copy of the $Sp(2)$ group. The result is
then a Chern-Simons theory for the $OSp(2|p)\times OSp(2|q)$ group with $%
\mathcal{N}=p+q$.

The $OSp(2|p)$ algebra is spanned by the generators $J_a ^+$ of $Sp(2)$, $%
M_{ij}$ of $SO(p)$ and $Q_{\alpha} ^i$ satisfying
\begin{align}
[J_a ^+,M_{ij}]& =0\, ,  \notag \\
[J_a ^+,J_b ^+] & = \epsilon_{abc}J^c  \notag \\
[J_a ^+, Q_{\alpha} ^i ]& = -\frac{1}{2}(\Gamma _a)^{\beta} _{\alpha} Q^i
_{\beta}\, ,  \notag \\
[M_{ij} , M_{kl}]& = \eta_{li} M_{kj} - \eta_{ki}M_{lj} - \eta_{lj}M_{ki}
+\eta_{kj}M_{li}\, ,  \label{ospp} \\
[M_{ij}, Q^k _{\alpha}] &= (m_{ij})^k _{\ l} Q_{\alpha}^l \, ,  \notag \\
\{ Q_{\alpha} ^k , Q_{\beta} ^l\}& = (C\Gamma^a )_{\alpha \beta} J^+ _a
\eta^{kl} +\frac{1}{2}c_{\alpha \beta} M^{kl}\, ,  \notag
\end{align}
where $(m^{kl}) ^i _{\ j}$ are given by:
\begin{gather}
(m^{kl}) ^i _{\ j} = \left( \eta^{ik} \delta^l _{\ j} - \eta^{il} \delta ^k
_{\ j} \right)  \label{m}
\end{gather}
and provide a representation for SO($p$). We define a superconnection $%
\mathcal{A}_{\pm}$.
\begin{gather}
\mathcal{A}_{+} = A_{+}^a J_a + \frac{\psi ^{\alpha} _i}{\sqrt{l}} Q^i
_{\alpha} - \frac{1}{2} a^{kl}M_{kl}
\end{gather}
with the index $i$ running from $1$ to $p$. Similarly we define a
superconnection for the $OSp(2|q)$ algebra:
\begin{gather}
\mathcal{A}_{-} = A_{-}^a J_a + \frac{\psi ^{\alpha} _I}{\sqrt{l}} Q^I
_{\alpha} - \frac{1}{2} b^{IJ}M_{IJ}
\end{gather}
with the index $I$ running from $1$ to $q$.

The matrix representation for the connection $\mathcal{A}$ is given in
Appendix \ref{Appendix_pq} where the Chern-Simons term was derived. The
Chern-Simons Lagrangian for $OSp(2|p)$ is:
\begin{equation*}
L_{+}=\frac{1}{2}L_{CS}(A_{+})-L_{CS}(a_{\ j}^{i})-\bar{\psi}^{k}\nabla
_{+}\psi _{k}
\end{equation*}%
where the covariant derivative is
\begin{equation*}
\nabla _{+}\psi _{j}=d\psi _{j}-\frac{1}{2}A_{+}^{a}\Gamma _{a}\psi
_{j}-a_{\ j}^{k}\psi _{k}\,.
\end{equation*}%
For $OSp(2|q)$ the Lagrangian is:
\begin{equation*}
L_{-}=\frac{1}{2}L_{CS}(A_{-})-L_{CS}(b_{\ J}^{I})-\bar{\psi}^{I}\nabla
_{-}\psi _{I}
\end{equation*}%
The covariant derivative is
\begin{equation*}
\nabla _{-}\psi _{I}=d\psi _{I}-\frac{1}{2}A_{-}^{a}\Gamma _{a}\psi
_{I}-b_{\ I}^{J}\psi _{J}.
\end{equation*}%
Our starting point to construct the most general supergravity action in
three dimensions is to take a combination of the $L_{+}$ and $L_{-}$ actions
of the form
\begin{equation}
I=\kappa l(L_{+}-L_{-})+k(L_{+}+L_{-}).
\end{equation}%
With the standard identification $A_{\pm }^{a}=\omega ^{a}\pm e^{a}/l$, the
action $I$ can be rewritten as
\begin{gather}
I=I_{AdS}-l\left[ \kappa +\frac{k}{l}\right] \text{Tr}\left( ada+\frac{2}{3}%
a^{3}\right)   \notag \\
+l\left[ \kappa -\frac{k}{l}\right] \text{Tr}\left( bdb+\frac{2}{3}%
b^{3}\right)   \notag \\
-\left[ \kappa +\frac{k}{l}\right] \bar{\psi}^{i}\left( \left[ D-\frac{1}{2l}%
e^{a}\Gamma _{a}\right] \psi _{i}-a_{ij}\psi ^{j}\right)   \notag \\
+\left[ \kappa -\frac{k}{l}\right] \bar{\psi}^{I}\left( \left[ D+\frac{1}{2l}%
e^{a}\Gamma _{a}\right] \psi _{I}-b_{IJ}\psi ^{J}\right) \ ,
\end{gather}%
where the gravitational part of the action, $I_{AdS}$, is given by eq.(\ref%
{ADS_notorsion}). This is the known $(p,q)$ AdS supergravity action \cite%
{townsend}.

We now construct the more general AdS supergravity action. This is obtained
by applying the map $\omega ^{a}\rightarrow \omega ^{a}+\beta e^{a}$. The
action is then mapped
\begin{equation}
I\rightarrow I^{\prime }
\end{equation}%
with
\begin{equation*}
I^{\prime }=I_{G}+I_{a}+I_{b}+I_{\psi }^{+}+I_{\psi }^{-}\,,
\end{equation*}%
where $I_{G}$ is the most general gravity action in three dimensions given
by eq. (\ref{IG}) and the pieces involving the gauge fields are
\begin{equation*}
I_{a}:=l\alpha _{+}\text{Tr}\left( ada+\frac{2}{3}a^{3}\right) \,,
\end{equation*}%
\begin{equation*}
I_{b}:=l\alpha _{-}\text{Tr}\left( bdb+\frac{2}{3}b^{3}\right) \,,
\end{equation*}%
and the fermionic pieces are given by
\begin{equation*}
I_{\psi }^{+}:=\alpha _{+}\bar{\psi}^{i}\left( \left[ D-\frac{1}{2}\left(
\beta +\frac{1}{l}\right) e^{a}\Gamma _{a}\right] \psi _{i}-a_{ij}\psi
^{j}\right) \,,
\end{equation*}%
\begin{equation*}
I_{\psi }^{-}:=\alpha _{-}\bar{\psi}^{I}\left( \left[ D-\frac{1}{2}\left(
\beta -\frac{1}{l}\right) e^{a}\Gamma _{a}\right] \psi _{I}-b_{IJ}\psi
^{J}\right) .
\end{equation*}

The coefficients $\alpha_\pm$ are defined by:
\begin{gather}
\alpha_\pm:= \mp\left[\alpha_1 - \alpha_3 \left(\beta \mp\frac{1}{l}\right)%
\right]\, .
\end{gather}

The action is invariant under Lorentz transformations, AdS boosts and
supersymmetry as well as the $SO(p) \times SO(q)$ gauge symmetry.

Under a SUSY transformation the fields transform as

\begin{eqnarray}
\delta e^{a} &=&\frac{1}{2}\bar{\epsilon}^{i}\Gamma ^{a}\psi _{i}+\frac{1}{2}%
\bar{\epsilon}^{I}\Gamma ^{a}\psi _{I}\ ,  \notag \\
\delta \psi ^{i} &=&D\epsilon ^{i}-\frac{1}{2}\left( \beta +\frac{1}{l}%
\right) e^{a}\Gamma _{a}\epsilon ^{i}\ ,  \notag \\
\delta \psi ^{I} &=&D\epsilon ^{I}-\frac{1}{2}\left( \beta -\frac{1}{l}%
\right) e^{a}\Gamma _{a}\epsilon ^{I}\ , \\
\delta \omega ^{a} &=&-\frac{1}{2}\left( \beta -\frac{1}{l}\right) \bar{%
\epsilon}^{i}\Gamma ^{a}\psi _{i}-\frac{1}{2}\left( \beta +\frac{1}{l}%
\right) \bar{\epsilon}^{I}\Gamma ^{a}\psi _{I}\ ,  \notag \\
\delta a_{\ j}^{i} &=&-\frac{1}{2\sqrt{l}}(\bar{\epsilon}^{i}\psi _{j}-\bar{%
\psi}^{i}\epsilon _{j})  \notag \\
\delta b_{\ J}^{I} &=&-\frac{1}{2\sqrt{l}}(\bar{\epsilon}^{I}\psi _{J}-\bar{%
\psi}^{I}\epsilon _{J})  \notag
\end{eqnarray}

In the case of more than one supersymmetry one cannot take naively the limit
$l\rightarrow \infty $ to obtain the Poincar\'{e} case as it does not give a
supersymmetric theory. The correct approach is shown in ref. \cite{HIPT}
where a direct sum of the standard $(p,q)$ super AdS algebra and an $%
so(p)\oplus so(q)$ algebra is considered. With a suitable change of basis
one obtains the correct Poincar\'{e} limit. Therefore in order to obtain the
most general Poincar\'{e} supergravity theory one can apply the map to the
construction described in ref. \cite{HIPT}.

\section{Conclusions and discussion}

We have constructed the locally supersymmetric extension of the most general
gravity theory in three dimensions leading to first order field equations
for the vielbein and the spin connection. We have seen that the construction
of this theory can be simplified by making use of a mapping between
different theories of gravity described in section \ref{Map}. This mapping
is obtained by a shift in the spin connection of the form $\omega
^{a}\rightarrow \omega ^{a}+\beta e^{a}$. Notice that it is a special
feature of three dimensions that the spin connection can be dualized to have
one Lorentz index.

It is worth to point out that the mapping is useful also for finding
solutions. For example, it is known that the black hole solution with
torsion \cite{Garcia-Hehl} has the same metric as the BTZ black hole but the
spin connection is given by $\omega^a = \tilde{\omega}^a -\beta e^a$, where $%
\tilde\omega^a$ is the Riemannian connection. From our discussion, this
result can be seen immediately. We know that the theory that gives torsion $%
-\beta$ can be obtained from the torsion-free theory by replacing $%
\tilde\omega^a$ with $\omega^a + \beta e^a$ everywhere in the action. This
means that this mapping is also a mapping between solutions of the two
theories. Given that the BTZ is a solution of the torsionless theory, the
corresponding solution of the new theory is simply $\omega = \tilde\omega -
\beta e$.

It is also interesting to study the black hole solution with torsion for its
thermodynamical properties. In fact it was shown \cite%
{Blagojevic:2006jk,BC-JHEP} that the black hole entropy acquires a
correction to the usual Bekenstein-Hawking formula proportional to the inner
horizon. Also the conserved charges of gravity with torsion have been
studied \cite{Blagojevic:2005hd}. An interesting application is given by the
computation of the mass and angular momentum of the BTZ black hole with
torsion. It is found that there is a mixing between the integration
constants M and J which in the torsion free case are respectively the mass
and angular momentum. It should be possible to obtain the same results by
use of the mapping.

It would also be interesting to find the BPS states and the asymptotic
dynamics of the here proposed supergravity model with torsion. By virtue of
the map presented here, one naturally expects that the boundary conditions
should generally be different for two different theories related by the map,
so that the physics of both theories is clearly inequivalent.

\textit{Note added:} After the completion of this paper, the structure of
asymptotic symmetries for the supergravity theory presented in Section $\ref%
{pq_section}$ for the case of $\mathcal{N}=1+1$ was studied in Ref. \cite%
{BC-Last}, where it was found that the asymptotic Poisson bracket algebra of
the canonical generators has the form of two independent super-Virasoro
algebras with different central charges.

\paragraph{Acknowledgements:}

This work was partially funded by FONDECYT grants 1040921, 1051056,
1061291  3060016 and 3070057. The generous support to CECS by
Empresas CMPC is also acknowledged. CECS is a Millennium Science
Institute and is funded in part by grants from Fundaci\'{o}n Andes
and the Tinker Foundation.

\begin{appendix}

\section{Notation and conventions}

We use the metric with the signature $(-,+,+)$. The Levi-Civita
symbol $\epsilon_{abc}$ is totally antisymmetric with
$\epsilon_{012}=1$, $\epsilon^{012} = -1$ and it satisfies the
following identities:
\begin{gather*}
 \epsilon_{abc}\epsilon^{ade}
 = -(\delta^d_b \delta^e_c - \delta^e_b \delta^d_c)\, ,
\qquad\epsilon_{abc}\epsilon^{abd}
 = -2\delta^d_c\, .
\end{gather*}

The Dirac matrices  satisfy the Clifford algebra $\{ \Gamma_a,
\Gamma_b\} = 2 \eta_{ab}$.  We choose
\begin{equation*}
\Gamma_0 = i \sigma_x, \quad \Gamma_1 = \sigma_y, \quad \Gamma_2 \sigma_z
\end{equation*}
where the $\sigma$'s are the usual Pauli matrices. Then we have the
identity
\begin{equation*}
\Gamma_{ab} = - \epsilon_{abc}\Gamma^c.
\end{equation*}
For a spinor $\epsilon^\alpha$ we define the Majorana conjugate $%
\bar\epsilon_\beta := C_{\beta\alpha}\epsilon^\alpha$, where $C$ is
the charge conjugation matrix satisfying $C^T = -C$ and
$(C\Gamma_a)^T = C\Gamma_a$.

\section{Matrix representation of $OSP(2|1)$ and construction of the
 $\mathcal{N} = 1 $ Lagrangian}\label{Appendix_N1}

We choose a set of matrices that form a representation of the
$OSp(2|1)$ Lie algebra (\ref{osp1}) given by:

\begin{gather}
J_a = \left(%
\begin{array}{cc}
-\frac{1}{2}(\Gamma_a)^\alpha_\beta & 0 \\
0 & 0%
\end{array}%
\right),\quad Q_\sigma = \left(%
\begin{array}{cc}
0 & \frac{1}{\sqrt{2}} \delta^\alpha_\sigma \\
-\frac{1}{\sqrt{2}}C_{\sigma\beta} & 0
\end{array}
\right).
\end{gather}

The $OSp(2|1)$ connection defined as $ \mathcal{A} = A^a J_a +
\frac{\psi^{\alpha}}{\sqrt{l}}Q_{\alpha}$, in this representation
takes the form
\begin{gather}
\mathcal{A} = \left(%
\begin{array}{cc}
-\frac{1}{2}A^a \Gamma_a & \frac{1}{\sqrt{2l}} \psi \\
\frac{1}{\sqrt{2l}}\bar{\psi} & 0
\end{array}
\right)
\end{gather}
Hence we can compute the curvature  defined as $ \mathcal{F}=
d\mathcal{A} + \mathcal{A}^2$ reads
\begin{gather}
\mathcal{F} = \left(%
\begin{array}{cc}
-\frac{1}{2}F^a \Gamma_a + \frac{1}{2 l} \psi\bar{\psi} & \frac{1}{\sqrt{2 l}%
}\nabla \psi \\
\frac{1}{\sqrt{2l}}\nabla\bar{\psi} & 0
\end{array}
\right)
\end{gather}
 Here $\nabla$ is the covariant derivative is given by $%
\nabla \psi = d\psi -\frac{1}{2} A^a \Gamma_a \psi$. We can
 use the fact that for Chern-Simons
thoeries we have $dL = \langle\mathcal{F}^2\rangle$, where the
invariant four-form is the supertrace of $\mathcal{F}^2$ to write
the Lagrangian in terms of the components of $\mathcal{A}$.
\begin{align}
\mathrm{STr}(\mathcal{F}^2) & = \frac{1}{2} F^a F_a -\frac{1}{2} F^a \bar{%
\psi} \Gamma_a \psi -\frac{1}{l}\nabla \bar{\psi} \nabla \psi \\
& = d\left(\frac{1}{2} L_\text{cs}(A) - \frac{1}{l}
\bar{\psi}\nabla \psi\right)= dL.
\end{align}

\section{Matrix representation for $OSp(2|p)$ and the construction
of the $\mathcal{N}=p+q$ supergravity
Lagrangian}\label{Appendix_pq}

We have the following matrix representation of the $OSp(2|p)$ Lie
algebra (\ref{ospp}):
\begin{gather}
J_a = \left(
\begin{array}{cc}
-\frac{1}{2} (\Gamma_a)_{\alpha \beta} & 0 \\
0 & 0%
\end{array}
\right) \\
M_{kl} = \left(
\begin{array}{cc}
0 & 0 \\
0 & -(m_{kl}) ^i _{\ j}%
\end{array}
\right) \\
Q_{\gamma} ^k = \left(
\begin{array}{cc}
0 & \frac{1}{\sqrt{2}} \delta^{\alpha} _{\gamma} \delta^k _j \\
-\frac{1}{\sqrt{2}}C_{\gamma \beta} \eta^{ki} & 0%
\end{array}
\right)
\end{gather}
Where $(m^{kl}) ^i _{\ j}$ is defined as in eq.(\ref{m})

The matrix representation for the connection $\mathcal{A}= A^a J_a +
\frac{\psi ^{\alpha}}{\sqrt{l}} Q_{\alpha} -
\frac{1}{2}a^{kl}M_{kl}$ is
\begin{gather}
\mathcal{A}= \left(
\begin{array}{cc}
-\frac{1}{2}A^a\Gamma_a & \frac{1}{\sqrt{2l}}\psi_j \\
\frac{1}{\sqrt{2l}}\bar{\psi}^i & a^i_{\ j}%
\end{array}
\right)
\end{gather}
So the field strength $\mathcal{F} = d\mathcal{A} +\mathcal{A}^2$
becomes
\begin{gather}
\mathcal{F} = \left(
\begin{array}{cc}
F +\frac{1}{2l} \psi_k \bar{\psi}^k & \frac{1}{\sqrt{2l}}\nabla \psi_j \\
\frac{1}{\sqrt{2l}} \nabla \bar{\psi}^i & f^i_{\ j} +\frac{1}{2l}
\bar{\psi}^i
\psi_j%
\end{array}
\right)
\end{gather}
where we have defined the field strenght of the $SO(P)$ gauge field,
$f^i_{\ j} =da^i_{\ j} + a^i_{\ k} a^k _{\ j}$. We obtain then
\begin{gather}
Str[\hat{F}^2]= \frac{1}{2}F^aF_a - f^k _{\ j} f^j _{\ k} -
d(\bar{\psi}^k \nabla\psi_k )= dL
\end{gather}
So we obtain the Lagrangian
\begin{gather}
L = \frac{1}{2}L_{CS}(A) - L_{CS}(a^i _{\ j}) - \bar{\psi}^k\nabla
\psi_k\ .
\end{gather}

\end{appendix}

\end{document}